\documentclass[journal]{IEEEtran}
\ifCLASSINFOpdf
  \usepackage[pdftex]{graphicx}
\else
  \usepackage[dvips]{graphicx}
\fi
\usepackage{array}

\usepackage{amsmath}
\usepackage{mathtools}

\usepackage{booktabs}
\usepackage{cite}

\usepackage{gensymb}

\begin{document}
%
\title{High-$\mathrm{Q}$ Whispering-Gallery-Modes Microresonators in the Near-Ultraviolet Spectral Range}
%
%
%

\author{Georges~Perin, Yannick~Dumeige, Patrice~Féron, and Stéphane Trebaol
\thanks{Univ. Rennes, CNRS, Institut Foton - UMR 6082, F-22305 Lannion, France. e-mail: stephane.trebaol@enssat.fr}

}

%
%

\markboth{Journal of \LaTeX\ Class Files,~Vol.~14, No.~8, August~2015}%
{Shell \MakeLowercase{\textit{et al.}}: Bare Demo of IEEEtran.cls for IEEE Journals}
%



\maketitle

\begin{abstract}
High-$Q$ whispering gallery mode microresonators are cornerstones to develop compact coherent sources like single frequency lasers or frequency combs. This work reports a detailed characterization of a whispering gallery mode microsphere in the near ultraviolet. Light coupling in the resonator  is obtain thanks to a robust angle-polished fiber allowing to investigate the different coupling regimes. An intrinsic $Q_0$-factor of $2.2 \times 10^8$ and a finesse of $7.3 \times 10^4$ are reported at a wavelength of 420 nm. Physical mechanisms contributing to the $Q_0$-factor are discussed and routes to improve the performances are drawn. Such high-$Q$ factor and high finesse are key ingredients to ease the study of photonic devices based on WGM microresonators. 
\end{abstract}

\begin{IEEEkeywords}
Whispering gallery modes, microsphere, near ultraviolet, high $Q$-factor
\end{IEEEkeywords}

%
\IEEEpeerreviewmaketitle

\section{Introduction}
%
%
%
%
\IEEEPARstart{W}{hispering} gallery mode (WGM) microresonators have been extensively studied since decades for their unique properties like large $Q$-factor, long photon lifetime storage and small optical volume. Those unique features allow to demonstrate linear, non-linear and quantum photonic devices based on WGM resonators. Applications range from narrow-band filter \cite{Collodo2014}, compact optical memory \cite{Huet2016}, optical locking for laser linewidth narrowing \cite{Liang2015}, Raman microlaser \cite{Spillane2002},  on-chip frequency comb generation  \cite{DelHaye2007}, low threshold lasing \cite{Sandoghdar1996}, biological sensing \cite{Ward2011} or quantum electrodynamics experiments \cite{Buck2003, Junge2013}. Various shape of microresonators have been studied \cite{Ward2011} as microspheres, millidisks, microdisks and microtoroids in various materials like fused silica \cite{Chiasera2010}, fluoride and chalcogenide glasses or cristals \cite{Savchenkov2019,Lin2014, Elliott2010}. Very high $Q$-factors have been reported up to $10^{10}$ in passive fused silica microsphere\cite{Gorodetsky1996} and a record level of $3\times10^{11}$ in CaF$_2$ crystalline millidisk resonator \cite{Savchenkov2007b}. Despite the huge interest for those compact microresonators, they have been mainly studied in the near infrared region.\\
Since few years, the need of short wavelength compact coherent sources like single-frequency lasers and optical frequency combs, as tools for high precision spectroscopy \cite{Ludlow2015, Maurice2020}, pushes the research community to investigate the potential of WGM resonators at short wavelengths \cite{Donvalkar2018, Savchenkov2019, Savchenkov2020, Lee2017, Lin2012}. Until now, only millidisk configurations have been investigated with record $Q$-factors estimated to be as high as $10^9$ at 370 nm \cite{Savchenkov2019}. However, microsphere resonator shape can be of interest in particular for QED experiment where high-$Q$ factor should be associated with small cavity volume \cite{Buck2003}.\\
In this work, we report the full characterization of fused silica WGM microspheres in the near ultraviolet at 420 nm. Thanks to the cavity ringdown method~\cite{Dumeige2008}, we demonstrate an intrinsic microsphere $Q$-factor in excess to $10^8$ limited by surface roughness. Robustness of the coupling setup based on angle-polished fiber offer the possibility to investigate the different coupling regimes. Critical coupling regime is obtained for a gap distance of 40 nm with a contrast of 100 $\%$.  In the near infrared region, water deposition on the microsphere surface degrades the $Q$-factor over time toward the $10^7-10^8$ range \cite{Gorodetsky1996, Vernooy1998}. Thanks to the water transparency window around 420 nm, no degradation is observed, which allow stable operation of WGM based photonic devices at short wavelengths.\\
In section \ref{sec:fabrication}, we report the fabrication process of the microspheres, in section \ref{sec:coupling} the coupling bench based on an angle-polished fiber is described. Microsphere characterization is investigated in section \ref{sec:characterization}, in particular, a description of the  characterization technique that allows $Q$-factors and coupling regime determination is given. Finally, in section \ref{sec:limitations} we discuss the mechanisms that contribute to the intrinsic $Q$-factor and identify the surface roughness as the main limitation.
\begin{figure}[!t]
\centering
\includegraphics[scale=0.5]{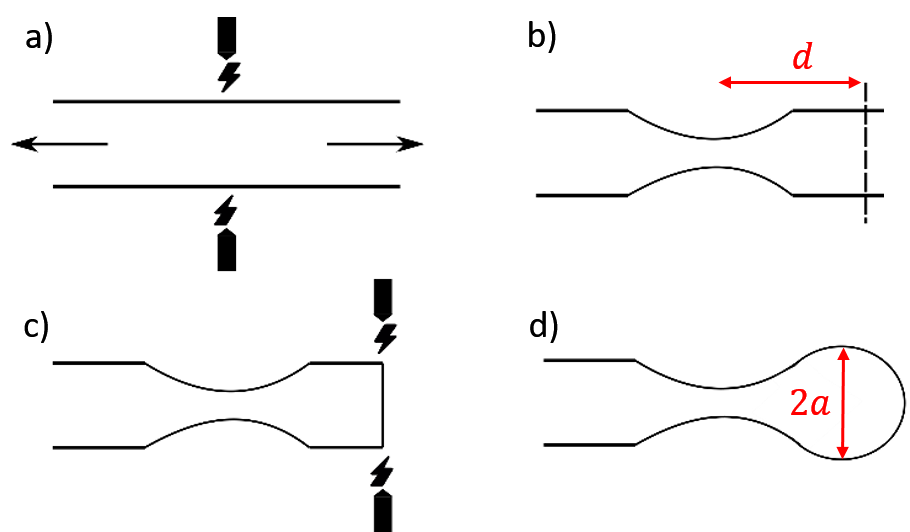} 
\caption{Microspheres are fabricated in 3 steps by use of a fiber splicer. The heat source is brought by two electrodes placed on either side of the fiber. a) Fiber elongation give rise to a fiber taper, b) the fiber is cleaved at a distance $d$ from the taper region, to determine the quantity of silica to melt, c) the fiber tip is melted to form the microsphere, d) schematic of the final microsphere sample. $a$ is the sphere radius.}
\label{fig_fab_steps}
\end{figure}
\section{Microsphere fabrication}\label{sec:fabrication}
Spherical microresonators are obtained by melting the tip of a pure silica fiber  with a diameter of 125 $\mu$m (F300 from Heraeus). The fiber presents a low rate of hydroxide ions ($< 0.2 $ppm) responsible in light absorption that degrade the transparency of the material. It turns that propagation losses in the fiber are given to be 30 dB/km around 420 nm. A fiber splicer is used to fabricate the microresonators. The three fabrication steps are shown on Figure \ref{fig_fab_steps}. The heat source is bring by two electrodes that produce an electrical arc \cite{Inga2020}. First, the fiber is elongated to form a taper with a central diameter below 50 $\mu$m and a narrowed region of around 700 $\mu$m. Then the fiber is cleaved at a distance $d$ that sets the silica quantity brought for the sphere fabrication. The fiber diameter can be adjusted by varying $d$ ($d\approx 1$ mm for a microsphere radius $a\approx 150~\mu$m) . Finally, the electrode driving current is appropriately adjusted to reach the glass transition temperature. The surface tension forces bring naturally a spheroid shape to the microresonator. Microsphere diameters ranging from 140 to 400 µm can be fabricated using our method. Inset of Figure \ref{fig_experimentschematic} show a picture of the 275 $\mu$m diameter microsphere used in this work.

\section{Angle-polished fiber coupling}\label{sec:coupling}
\begin{figure}[!t]
\centering
\includegraphics[scale=0.75]{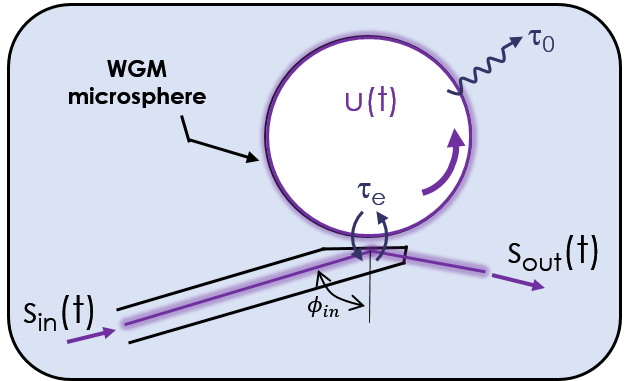} 
\caption{Sketch of a WGM microsphere coupled to an angle-polished fiber pigtail. The input and output field are $s_\text{in}(t)$ and $s_\text{out}(t)$ respectively. $u(t)$ is the resonator mode amplitude, $\tau_0$, the intrinsic photon lifetime and $\tau_e$ the coupling photon lifetime. $\phi_\text{in}$ stands for the optimal cutting angle for phase matching condition.}
\label{fig_schematicWGM}
\end{figure}

\begin{figure}[!t]
\centering
\includegraphics[scale=0.6]{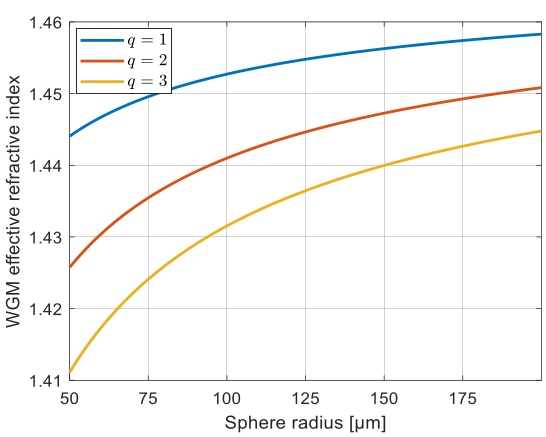} 
\caption{Evolution of the effective refractive index $n_\text{sphere}$ as a function of the microsphere radius $a$. The Airy function root values are $2.338$, $4.088$ and $5.521$ for $q = 1,2,3$ respectively. The following parameter values are used : operating wavelength $\lambda=420$ nm, silica optical index $n= 1.4681$, angular momentum $l$ is obtained from $l=2\pi a n/\lambda$.}
\label{fig_effective_index}
\end{figure}

Light in a whispering-gallery mode microresonator propagates along the surface by total internal reflection implying a strong light confinement and high-$Q$ factor. Reciprocally, the difficulty lies in an efficient light coupling inside the cavity. The solution consists in overlapping the WGM tail and the input beam evanescent field. To this end, three coupling systems are usually reported : prism~\cite{Braginsky1989, Gorodetsky1999}, tapered fiber~\cite{Knight1997} and angle-polished fiber~\cite{Ilchenko1999}. In the present study, the angle-polished fiber have been implemented since it offers a good compromise in term of coupling efficiency, ease of implementation and robustness especially at short wavelengths where working distance between the microsphere and the coupling setup consists in few tens of nm. A schematic of the angle-polished fiber is shown on Figure \ref{fig_schematicWGM}. The light coupled inside the fiber core experiences a total internal reflection upon the angled surface. The evanescent part of the field is subsequently coupled into the microsphere. To obtain the phase matching between the waveguided mode and the considered microsphere whispering gallery mode, one have to fulfill the following phase matching condition~\cite{Ilchenko1999} :
\begin{equation}
\phi_\text{in} = \sin^{-1}\left(\frac{n_\text{sphere}}{n_\text{fiber}} \right)
\end{equation}
where $n_\text{sphere}$ is the effective refractive index of the WGM and $n_\text{fiber}$ is the single-mode-fiber-effective refractive index. To determine the effective index for WGM modes inside the microsphere, we first have to calculate the resonant frequency of TE$_{l,m,q}$ using the following expressions \cite{Lam1992, Ilchenko1999}:
\begin{equation}
\omega_{l,q} = \frac{c}{na}\left[\nu +2^{-1/3}\alpha_q\nu^{1/3}-\frac{n}{\sqrt{n^2-1}}\right]
\end{equation}
with $l$, $m$ and $q$ the polar, azimuthal and radial numbers of the electromagnetic field \cite{Chiasera2010}, $c$ is the speed of light, $n$, the refractive index of silica ($n=1.4681$ at 420 nm \cite{Malitson1965}), $a$ the microsphere radius, $\nu=l+\frac{1}{2}$, $\alpha_q$ is the $q_\text{th}$ root of the Airy function $Ai(-z)$. The effective refractive index of the WGM is obtained from $n_\text{sphere}=cl/a\omega_{l,q}$. Figure \ref{fig_effective_index} shows the value of $n_\text{sphere}$ as a function of the microsphere radius for the fundamental and the two first high order WGMs ($q=1,2,3$). For a microsphere diameter of 275 µm, the effective refractive index is $n_\text{sphere}=1.4556 (1.4660)$ for $q=1 (2)$.\\
The effective fiber index, $n_\text{fiber}$ is experimentally determined to be 1.4675, which finally gives an angle-polished fiber $\phi_\text{IN}= 82.7 (80.2) \degree$ for $q=1 (2)$. We finally fabricate an angle-polished fiber with $\phi_\text{IN}= 81 \pm 1 \degree$ to couple light inside the microsphere.
Considering TM modes for the angle-polished fiber design gives a similar value for $\phi_\text{IN}$.\\
To collect the signal extracted from the microsphere the tip of the fiber is cleaved perpendicularly to the light beam after reflection at the glass/air interface (see Figure \ref{fig_schematicWGM}). The gap between the fiber and the microsphere can be finely tune from 10 nm to few µm by step of 10 nm. 
\section{Microsphere characterization}\label{sec:characterization}
To characterize the microsphere and in particular to determine the quality factor of the resonator, usual laser scanning technique \cite{Carmon2004} cannot be used on high-Q ($Q> 10^7$) resonator. Indeed, thermal effect induced by material light absorption strongly distorted the resonance \cite{Carmon2004} especially at short wavelengths where the linear absorption coefficient is 100 times larger than at telecom wavelengths. However, the cavity ringdown technique have been extensively used to determine the linear and nonlinear properties of high-Q resonator \cite{Dumeige2008, Conti2011,Henriet2015, Rasoloniaina2014}. This technique allows to accurately estimate the intrinsic linear and nonlinear losses, the coupling regime and the linear and nonlinear dispersive properties of WGM resonators \cite{Rasoloniaina2015}. We propose in this paper to applied this technique to microspheres characterization close the near UV range at 420 nm.
\subsection{System description}\label{subsec:system}
The fiber-coupled-microsphere experimental arrangement is depicted in Figure \ref{fig_schematicWGM}. The evanescent tail of the single-mode fiber is coupled into the probe WGM mode whose amplitude is $u(t)$. The coupling photon lifetime is expressed by $\tau_e$ and the intrinsic photon lifetime is denoted by $\tau_0$. The later represent the microsphere losses experienced by the mode $u(t)$. The total lifetime is then deduced $\tau^{-1}=\tau_0^{-1}+\tau_e^{-1}$ and the loaded Q factor is expressed as $Q=\omega_0 \tau/2$ where $\omega_0$ is the angular frequency of the probed WGM. In the same way, the intrinsic Q factor is given by $Q_0 = \omega_0 \tau_0/2$.\\
For such high-finesse microresonator ($\mathcal{F}> 10^4$) the collected output signal $s_\text{out}(t)$ can be obtained from the input signal amplitude $s_\text{in}(t)$ by solving the following system \cite{Dumeige2008}:
\begin{equation}\displaystyle \label{system}
\begin{cases}
\dfrac{du}{dt}=\left( j\left[\omega_0 - \frac{1}{\tau}\right]\right)u(t)+\sqrt{\frac{2}{\tau_e}}s_\text{in}(t) \\
s_\text{out}(t)=-s_\text{in}+\sqrt{\frac{2}{\tau_e}}u(t)
\end{cases}
\end{equation}
The instantaneous amplitude of the whispering gallery mode $u(t)$ can be determine by integration of the equation system (\ref{system}). The shape of the amplitude $u(t)$ will strongly depend on the shape of the input signal $s_\text{in}(t)$. In this work we consider frequency chirped input signal which means that the carrier frequency is linearly swept in time such as $s_\text{in}(t)=s_0e^{j\phi(t)}$ with $\phi(t)=\omega_i t+\pi V_St^2$ where $s_0$ and $\omega_i$ are the amplitude and the initial angular frequency of the input signal respectively. $V_S$ is the frequency scanning speed of the probe laser. 
\subsection{Cavity ringdown experimental setup}
\begin{figure}[!t]
\centering
\includegraphics[scale=0.75]{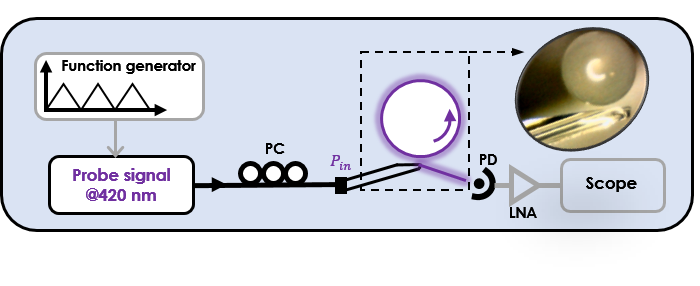} 
\caption{Experimental setup for WGM  microsphere characterization. PC, PD, and LNA stand for polarization controller, photodiode and low noise electrical amplifier respectively. $P_\text{in}$ is the probe laser input power fixed at 4 mW. Inset: 275 µm diameter fiber coupled to a angle-polished fiber.}
\label{fig_experimentschematic}
\end{figure}
The experimental setup for the characterization of the resonators is presented in Figure \ref{fig_experimentschematic}. A single-mode laser source whose emission is centered around 420 nm and continuously tunable over 20 GHz is used to scan the resonances of the microresonator under test. The laser signal is injected, using the angle-polished fiber, towards the resonator. The resonator is passively isolated on a stabilized table placed in a box in order to isolate it from environmental perturbations. The angle-polished fiber is inserted in a rotating stage allowing roll and pitch in addition to the x, y, z translations. At the output of the coupling fiber, light collection is carried out using a fast photodiode with a bandwidth of 2 GHz followed up by a low noise electrical amplifier wired to an oscilloscope.\\
The cavity ringdown technique consists in rapidly sweeping the frequency of the probe laser across a resonance of the cavity, i.e. faster than the relaxation time $\tau$ of the resonator, this leads to interference between the signal extracted from the cavity and the signal transmitted through the angle-polished fiber. Measured signal consists of oscillations with an exponentially decaying envelope as shown in Figure \ref{fig_CRDM} a) (black curve).

\subsection{$Q$-factors estimation}
\begin{figure}[!t]
\centering
\includegraphics[scale=0.85]{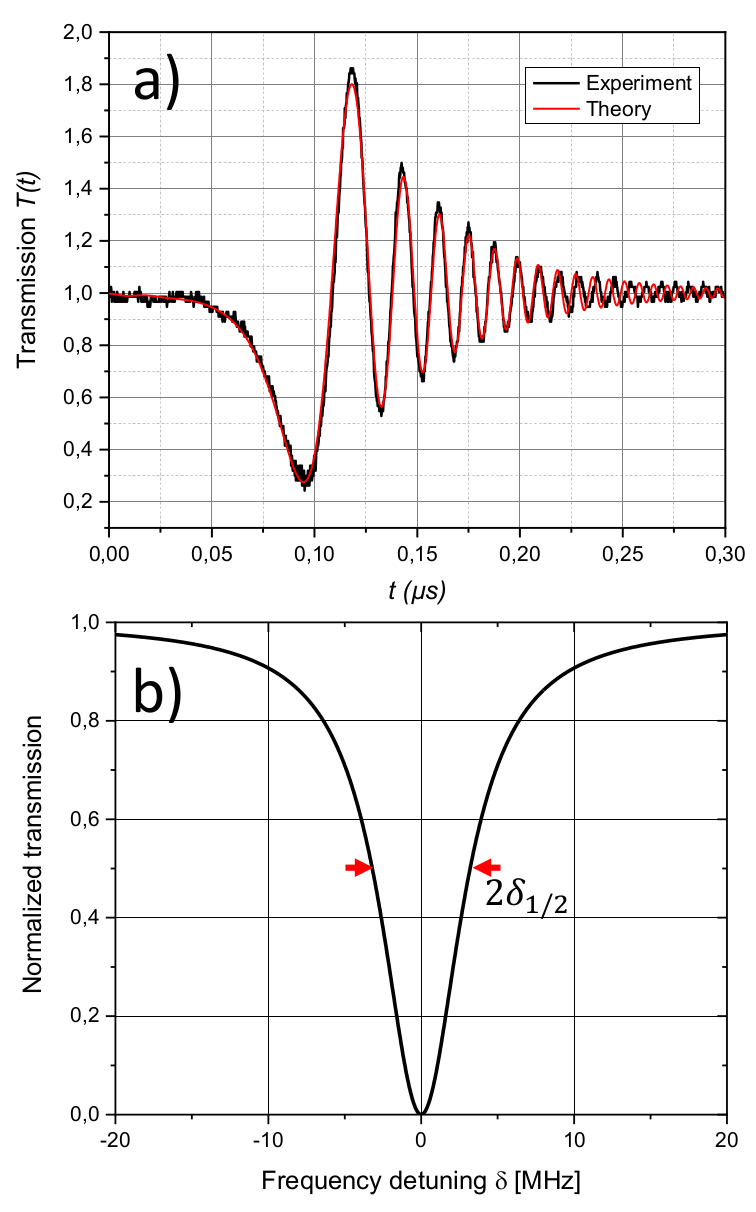} 
\caption{a) Transient response of WGM microsphere measurement obtained by cavity ringdown technique. Extracted parameters from the fit are $\tau_0=98$ ns, $\tau_e=101$ ns, and $V_s=840$ MHz$/\mu$s. b) Calculated Lorentzian shape transmission of the WGM microsphere from measured $\tau_0$ and $\tau_e$ using Eq.(\ref{eq:Transmission}). The FWHM, $2\delta_{1/2}$, is evaluated to $6.5$ MHz.}
\label{fig_CRDM}
\end{figure}

To estimate quality factors $Q$ and $Q_0$ from the transmission signal, $\tau_0$ and $\tau_e$ lifetimes are extracted by fitting the experimental curve with the theoretical model described above (see sec. \ref{subsec:system}). The fitting procedure is given in previous works \cite{Dumeige2008,Rasoloniaina2014, Rasoloniaina2015}. The fitting parameters are the lifetimes $\tau_0$ and $\tau_e$, and the frequency scanning speed $V_S$. The fitting curve is displayed in red color on Figure \ref{fig_CRDM} a). A good overlap with the experimental curve is observed. Extracted parameters give $\tau_0 = 98 \pm 10 $ ns, $\tau_e = 101 \pm 10$ ns, which gives a loaded $Q$-factor of $Q= (1.1\pm~0.4)\times 10^8$ and an intrinsic $Q$- factor of $Q_0=(2.2\pm~0.4)\times 10^8$.
Finally, thanks the CRDM technique, linear parameters like $\tau_0$ and $\tau_e$ can be evaluated. Then resonator transmission in the linear regime can be determined from \cite{Dumeige2008}:
\begin{equation}\label{eq:Transmission}
T(\delta)=\frac{(1/\tau_e-1/\tau_0)^2+4\pi^2\delta^2}{(1/\tau_e+1/\tau_0)^2+4\pi^2\delta^2}
\end{equation}
with $\delta$ the frequency detuning from the resonance. The Lorentzian shape transmission spectrum is shown on Figure \ref{fig_CRDM} b). The full width at half-maximum (FWHM), $2\delta_{1/2}$ is related to the overall $Q$-factor by:
\begin{equation}
Q=\frac{\omega_0}{4\pi\delta_{1/2}}
\end{equation} 
which gives a FWHM, $2\delta_{1/2}= 6.5 \pm 2.4$ MHz.\\ 
\subsection{Coupling regime versus coupling strength}
\begin{figure}[!t]
\centering
\includegraphics[scale=0.65]{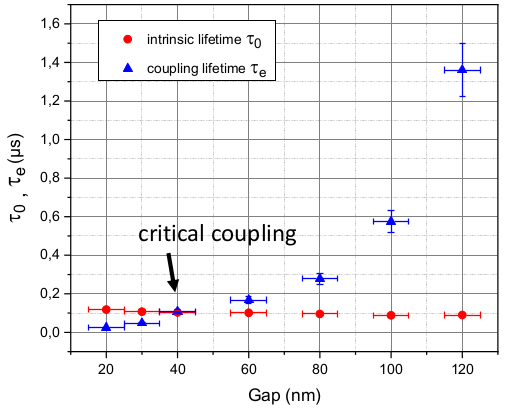} 
\caption{Intrinsic and coupling lifetimes evolution versus the gap between the angle-polished fiber and the microsphere.}
\label{fig_gap}
\end{figure}

The versatility of the angle-polished fiber allows the gap between the sphere and the coupling fiber to be finely tune by step of 10 nm. This offer the possibility to vary the coupling regime from undercoupling $\tau_e> \tau_0$ to overcoupling $\tau_e<\tau_0$ through  the critical coupling regime $\tau_e=\tau_0$.\\
Figure \ref{fig_gap} displays $\tau_0$ and $\tau_e$ values for gap variations from 20 to 120 nm. First we can notice that $\tau_0 = 100 \pm 10$ ns is independent from gap variations. This was expected since the intrinsic photon lifetime refers to the optical losses inside the microsphere and then should not be affected by the coupling strength. The intrinsic $Q$-factor can then be estimated to be $(2.2\pm0.4)\times 10^8$ and the finesse to $\mathcal{F}=\lambda Q_0/(n \pi 2 a)=(7.3 \pm 1.3) \times 10^4$. High finesse resonator highlight large intrinsic optical power. For example, an input power $P_\text{in}$ = 1 mW gives an intrinsic optical power of $P\approx P_\text{in} 2 \mathcal{F}/\pi\approx 46$ W. Resonators with high-$Q$ and high finesse are good candidates to exhibit nonlinear effects.\\
The critical coupling regime is of interest for a large range of linear and nonlinear applications. Here, our angle-polished fiber coupling bench allow to obtain this regime with a contrast close to $100 \%$ for a gap value of 40 nm demonstrating the potential of this coupling tool for further investigation of microsphere-based photonic devices in the near-UV.
\section{Limitation on the $Q$-factor in the near-UV}\label{sec:limitations}
\begin{figure}[!t]
\centering
\includegraphics[scale=0.40]{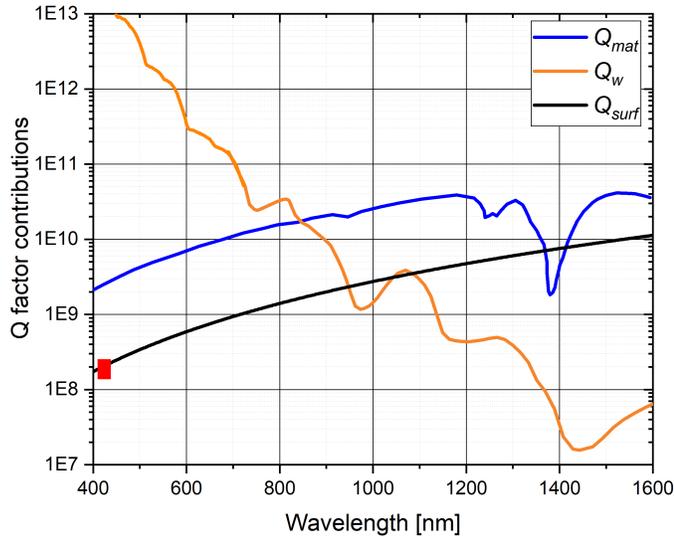} 
\caption{$Q$-factor contributions to the intrinsic $Q$-factor. $Q_\text{mat}$ is calculated from attenuation measurement on silica $F300$ by Humbach \emph{et al.}\cite{Humbach1996}. $Q_\text{w}$ is calculated from water absorption measurement from Pope \emph{et al.}~\cite{Pope1997} for the visible range and from Palmer \emph{et al.}~\cite{Palmer1974} for the near infrared range. Red square correspond to the measured intrinsic $Q$-factor in this work.}
\label{fig_Qversuslambda}
\end{figure}

The light field propagating inside the cavity suffers from losses caused by various mechanisms \cite{Gorodetsky1996}:
\begin{equation}
Q_0^{-1}=Q_\text{rad}^{-1} + Q_\text{mat}^{-1} + Q_\text{surf}^{-1} + Q_\text{w}^{-1}
\end{equation}
$Q_\text{rad}$ is related to the surface curvature, $Q_\text{mat}$  to the material losses, $Q_\text{surf}$ to the light scattering surface inhomogeneities and $Q_\text{w}$ to absorption losses from water deposition on the microsphere surface. The radiative losses are usually very small \cite{Gorodetsky1996} and do not contribute to the intrinsic Q factor. When the condition $2a/\lambda \geq 15$ is reached, the $Q_\text{rad}$ contribution to $Q_0$ is irrelevant, which is the case in the present study.\\ 
Light absorption in pure fused silica fiber is given to be $30$ dB/km at 420 nm \cite{Humbach1996} which gives :
\begin{equation}
Q_\text{mat}=\frac{2\pi n}{\alpha \lambda}\approx 3,1 \times 10^9
\end{equation}
As reported by Gorodestsky \emph{et al.} \cite{Gorodetsky1996}, water absorption on the surface of the microsphere can contribute to the intrinsic $Q$-factor limitation \cite{Vernooy1998}:
\begin{equation}
Q_\text{w}\approx \sqrt{\frac{\pi}{8n^3}}\frac{(2a)^{1/2}}{e\lambda^{1/2}\beta_w(\lambda)}
\end{equation}
where $e$ is the water layer thickness deposited on the microsphere surface and $\beta_w$ the water absorption coefficient. For $e=0.1$ nm \cite{Ganta2014} and $\beta_w=0.00454 $m$^{-1}$ at 420 nm \cite{Pope1997}, water absorption Q-factor is estimate to be $Q_w=2\times 10^{13}$ far above the measured $Q_0$ value. Here, we benefit from the water transparency window in the visible range. Contrary to reported studies in the NIR and (MIR) region, NUV fused silica microspheres do not suffer from water absorption that occurs minutes after the fabrication process.\\ 
The $Q$-factor related to surface scattering can be express as \cite{Gorodetsky2000,Lin2018}:
\begin{equation}\label{eq:Qsurf}
Q_\text{surf}\approx \frac{3\lambda^3 a}{8 n \pi^2 B^2 \sigma^2}
\end{equation}
Inga \emph{et al.} \cite{Inga2020} evaluated through AFM measurement the rms roughness to be $\sigma= 0.4 $ nm and a statistical correlation length $B=90.6$ nm for fused silica microsphere fabricated using electric arc as it is done in the present paper. This gives a $Q_\text{surf}=2.0\times 10^8$ at 420 nm, which is comparable to the measured intrinsic $Q$-factor. It is important to note that $Q_\text{surf}$ (see  Eq. (\ref{eq:Qsurf})) is inversely proportional to the square of $B$ and $\sigma$, and then strongly depends on this two parameters. Moreover, the method and the operation conditions for microsphere fabrication strongly affect measured values for $B$ and $\sigma$ \cite{Vernooy1998, Fan2000}. Figure \ref{fig_Qversuslambda} presents the contribution of the various $Q$-factors as a function of the wavelength for a given microsphere diameter of 275 µm. Physical limitations to the intrinsic $Q$-factor of fused silica microsphere strongly depend on the probing laser wavelength. Indeed, at telecom wavelength the main limitation is related to water absorption at the surface of the sphere that degrades the $Q$-factor to the $10^7 - 10^8$ range after minutes. Then, specific microsphere packaging should be arrange to maintain a dry environment. In the visible range, and in particular around 420 nm, water absorption is not an issue and the limiting effect is related to Rayleigh scattering due to surface inhomogeneities (see square box on Figure \ref{fig_Qversuslambda}). Surface quality similar to Ref. \cite{Vernooy1998} will push the $Q_\text{surf}$ above the $Q_\text{mat}$ value. Indeed, improvement in our microsphere fabrication process to reduce the surface imperfections should allow us to gain one order of magnitude on the $Q_0$ value.\\

\section{Conclusion}
We have presented a detailed characterization in the near UV of a whispering gallery mode microsphere system in which light is coupled using an angle-polished fiber. Thanks to a cavity ringdown technique, intrinsic losses and coupling strength are deduced. An intrinsic $Q_0$-factor of $2.2 \times 10^8$ and a finesse of $7.3 \times 10^4$ are then reported. A revue on the physical mechanisms contributing to the intrinsic $Q_0$ factor is proposed. Contrary, to WGM microspheres probed in the near infrared region, surface-water-deposition-induced light absorption is not the limiting contribution thanks to the water transparency window around 420 nm. This independence to water absorption excludes the need to maintain a dry environment through appropriate packaging for further device integration. Scattering onto surface inhomogeneities have been identified as the limiting mechanism to the $Q_0$ factor in the $10^8$ range. Improvement of the surface quality by optimizing the fabrication protocol should allow to gain one order of magnitude for the $Q$-factor. The method of fabrication and characterization given in this paper might open the way towards photonic devices based WGM microspheres in the near-UV range.


%

%
%

\section*{Acknowledgment}

The authors would like to thank Photonics Bretagne to provide the fiber sample for microsphere fabrication.\\
This work was supported by the ANR project COMBO (18-CE24-0003-01).

\ifCLASSOPTIONcaptionsoff
  \newpage
\fi



\bibliographystyle{IEEEtran}
\bibliography{thebibliography}
\end{document}